\documentclass[prl,twocolumn,showpacs,amsmath,amssymb]{revtex4}
\usepackage[]{graphicx}
\usepackage[]{epsfig}

\begin{document}

\title{Feshbach Resonances in Kerr Frequency Combs}

\author{Andrey B. Matsko and Lute Maleki}

\affiliation{OEwaves Inc., 465 N. Halstead Str., Pasadena, CA 91107}

\begin{abstract}
We show that both the power and repetition rate of a frequency comb generated in a nonlinear ring resonator, pumped with continuous wave (cw) coherent light, are modulated.  The modulation is brought about by the interaction of the cw background with optical pulses excited in the resonator, and occurs in resonators with nonzero high-order chromatic dispersion and wavelength-dependent quality factor. The modulation frequency corresponds to the detuning of the pump frequency from the eigenfrequency of the pumped mode in the resonator.
\end{abstract}


\maketitle

Nearly a century ago, the coupling of a bound state to a continuum was independently investigated by Herman Feshbach \cite{feshbach58ap,feshbach62ap} and Ugo Fano \cite{fano35nc,fano61pra,fano05jrnist} in different physical systems. This phenomenon is still widely studied as it allows tuning the scattering length in dilute atomic Bose-Einstein condensates \cite{timmermans99rp,chin10rmp}, and is involved in enhancement and suppression of scattering of electromagnetic waves in various physical systems \cite{miroshnichenko10rmp}. In this Letter we show theoretically that Feshbach resonances can be readily observed in a classical nonlinear optical system that involves spontaneous symmetry breaking and generation of solitary optical pulses from a continuous wave background.

A Feshbach resonance occurs when a bound molecular state energetically approaches a delocalized scattering state, so that even a weak coupling leads to strong mixing between these two states \cite{chin10rmp}. Similar phenomenon can be observed in optics, when a soliton propagating in a waveguide cavity interacts with dispersive waves. Here, a fundamental Schr\"odinger soliton can be considered as a particle \cite{kaup78prsla}, or a bound state, decoupled from the continuum, or a delocalized state. The soliton and the continuum do not interact in an ideal 1D lossless nonlinear waveguide with anomalous group velocity dispersion (GVD) \cite{haus93jqe,hausbook}. However, the introduction of gain, loss, frequency filtering, and nonlinear absorption in the mode-locked laser cavity initiates coupling between the cw background and a soliton, resulting in periodic changes of pulse parameters \cite{katz10ol,lee14prl}.

The coupling leads to a resonance in transfer function of pump power modulation to output power in the mode locked laser. The resonance occurs at a specific frequency  $f_{res}$ in the radio frequency (RF) beat between the soliton and the continuum field. The frequency can be found from expression $2\pi f_{res} T_R= \Phi$, where $T_R$ is the round trip time for the laser cavity and $\Phi(T_R)$ is the phase accumulated by the soliton envelope during a round trip in the cavity \cite{lee14prl}. The equality of the phase is similar to the equality of the energy of the bound and delocalized states required for a Feshbach resonance. A perturbation that breaks symmetry of the system is also needed to initiate energy exchange between the states and reveal the resonance. Frequency filtering and nonlinear absorption fulfill this function in the mode locked laser.

In this Letter we theoretically study Feshbach resonances in Kerr frequency combs. A nonlinear optical resonator pumped with cw light can produce an optical frequency (Kerr) comb \cite{delhaye07n,savchenkov08prl} and a train of ultrashort optical pulses having properties similar to Schr\"odinger solitons \cite{matsko12pra,saha13oe,herr14np}.  This is because the Kerr frequency comb formation process is phase matched if the GVD of the resonator modes is anomalous \cite{agha07pra,agha09oe,delhaye14prl}. Optical pulses are generated directly on a cw background in the nonlinear microresonator pumped with cw light.  Unlike the case of the mode locked laser, the interaction between continuum and a soliton is essential here, since the loss of  pulse energy is compensated by the nonlinear interaction of the pulses with the background. In turn, the loss of the background power is compensated by the external cw pump. Feshbach resonances reveal themselves as resonant enhancement of interaction of the cw background spectrum and the pulse characteristics.

The Feshbach resonance cannot be observed in an ideal nonlinear ring resonator since the Kerr frequency comb and off-resonant continuum are independent there. The resonant coherent cw background feeds the soliton. Nonresonant harmonics injected to the resonator along with the pump do not interact with the pulse. Consequently, the repetition rate of the pulse train leaving the resonator does not depend either on power or frequency of the cw optical pump \cite{matsko05pra,matsko13oe}. To observe the resonance, one needs to introduce a perturbation, which can be either high-order chromatic dispersion (e.g. third order dispersion, TOD) or a dependence of the mode bandwidth on the wavelength. In that case fluctuations of the optical continuum can be transferred to the repetition rate and the amplitude of the soliton; and the transfer is resonant. We found that in this case the beat frequency \cite{lee14prl}  is equal to the frequency detuning between the pump frequency, $\nu_p$, and the frequency of the pumped resonator mode, $\nu_0$:
\begin{equation} \label{res}
f_{res}=\nu_0-\nu_p.
\end{equation}
Interestingly, the beat frequency does not depend on the other parameters of the resonator and the pump laser, and stays the same for the fundamental (one pulse per round trip) and harmonic (many pulses per round trip) mode locking regimes. Therefore, the resonance enables measurement of the detuning of the pump light from the eigenfrequency of the resonator mode.

To describe Kerr frequency comb generation in a nonlinear ring resonator with a single coupling port we use Lugiato-Lefever \cite{matsko13oe,lugiato87prl,matsko09proc,coen13ol,balakireva13arch} equation
\begin{eqnarray} \label{basic1}
&& T_R\frac{\partial A}{\partial T} + \frac{i}{2}\beta_{2\Sigma}\frac{\partial^2
A}{\partial t^2} -i\gamma_{\Sigma}|A|^2A = \\ \nonumber &&
 - \left ( \alpha_{\Sigma}+\frac{T_c}{2}+i \delta_0 \right ) A+
i\sqrt{T_c P_{in}} e^{i\phi_{in}},
\\ && A_{out}=\sqrt{P_{in}} e^{i\phi_{in}}+i\sqrt{T_c} A, \label{basicout}
\end{eqnarray}
where $A(T,t)$ is the slowly varying envelope of the electric field inside resonator, $A_{out}(T,t)$ is the field leaving the resonator, $T$ is slow time and $t$ is the retarded time. By definition, time scale $T$ is much longer than the pulse round trip time $T_R=2\pi R/V_g$ ($V_g$ is the group velocity), $2\pi R$ is the length of the resonator circumference, $\alpha_{\Sigma}$ is the amplitude attenuation per round trip, $T_c/2$ is the coupling loss per one round trip (both $\alpha_\Sigma$ and $T_c$ are much less than unity), $\delta_0 = 2\pi(\nu_0-\nu_p)T_R$ is the normalized detuning of the pump light frequency from the frequency of the corresponding optical mode, $\gamma_{\Sigma}$ is the cubic nonlinearity, $\beta_{2\Sigma}$ is the GVD term.

Fundamental autosoliton solution of Eq.~(\ref{basic1}), found using variational method \cite{matsko13oe,hasegawa00jstqe}, is
\begin{eqnarray}\label{a}
A(T,t)&=&A_c+A_p(T,t), \;\;\;
A_c=\sqrt{P_c}e^{i \phi_c}, \\ \nonumber
A_p(T,t)&=&\left [ \frac{P_p(T)}{2 } \right ]^{1/2}
{\rm sech} \left [  \frac{t}{\tau(T)}\right ]
e^{i \Phi(T)},
\end{eqnarray}
where $P_c$ is the power of the DC background, $\phi_c$ is the phase of the background wave, $P_{p}=4\delta_0/\gamma_{\Sigma}$ is the pulse peak power,  $\tau$ is the pulse duration ($\tau^2=-2\beta_{2\Sigma}/(\gamma_{\Sigma} P_p)$), and $\Phi(T)$ is the phase of the pulse envelope. We find for the phase
\begin{equation}
\Phi(T_R)=\delta_0,
\end{equation}
which, in accordance with equation derived in \cite{lee14prl}, results in  Eq.~(\ref{res}). This means that the pulse excited in the microresonator resonantly interacts with frequency harmonics of the continuum separated from the pump frequency $\nu_p$ by $\nu_0-\nu_p$. However, solution of Eq.~(\ref{basic1}) show that the interaction does not occur in the ideal case \cite{matsko13oe}. The equation has to be perturbed by TOD or wavelength dependence of the coupling and loss coefficients of the material for the interaction to take place.

These perturbations are unavoidable in a ring resonators. For example, it was shown \cite{herr14np} that in a MgF$_2$ whispering gallery mode (WGM) resonator with free spectral range (FSR) of $35.2$~GHz, the spectrum is characterized by
\begin{equation} \label{freq}
\nu_l=\nu_0+FSR\times l+\frac 1 2 D_2 l^2+ \frac 1 6 D_3 l^3,
\end{equation}
where $D_2=16$~kHz is the parameter determined by GVD, and $D_3=-130$~Hz is the parameter determined by TOD of the resonator spectrum. The attenuation of  light in the resonator is also frequency dependent \cite{matsko09proc}. For instance, in the case of evanescent prism coupling, and Rayleigh scattering limited loss,
\begin{equation}
T_c(\nu) \sim \nu^{-3/2}, \;\;\; \alpha_{\Sigma}(\nu) \sim \nu^4.
\end{equation}
We performed a numerical simulation and found that even such a slow frequency dependence allows observation of the Feshbach resonance.

To reveal a Feshbach resonance in the Kerr comb we i) perturbed the ideal model described by Eq.~(\ref{basic1}) by introducing either TOD or wavelength dependence of resonator Q-factor and ii) introduced a small magnitude and relatively slow  modulation of the cw pump light. We discovered two well defined resonances by evaluating the dependence of the  comb repetition rate and comb power on the modulation frequency. One resonance corresponds to the relaxation oscillation of the frequency comb generator and the other is the Feshbach resonance predicted by the analytical model.

We solved a set of ordinary differential equations describing the behavior of 121 comb modes instead of integrating the Lugiato-Lefever equation. The set can be presented in a compact form as
\begin{equation} \label{set}
\dot{\hat a}_j=-2\pi( \gamma+i\nu_j)\hat a_j+ \frac{i}{\hbar} [\hat V,\hat a_j]+F_0(t) e^{-2 \pi i \nu t} \delta_{j0,j},
\end{equation}
where ${\hat a}_j$ is an annihilation operator for $j^{th}$ mode, $\delta_{j0,j}$ is the Kronecker's delta, $\hat V= -(\hbar g/2) (\hat e^\dag)^2 \hat e^2$, $\hat e= \sum \hat a_j$, $g=\hbar \omega_0^2 c n_2/({\cal V}n_0^2)$. To parameterize the problem we introduce a dimensionless pumping constant $f(t)=(F_0(t)/2\pi \gamma)(g/2\pi \gamma)^{1/2}$, where
\begin{equation}
F_0(t)=\sqrt{\frac{4 \pi \gamma P}{\hbar \omega_0}} [1+m \cos(2 \pi f_m t)]
\end{equation}
stands for the amplitude of the continuous wave modulated external pump, $\omega_0=2 \pi \nu_0$, $m$ and $f_m$ are modulation amplitude and frequency, respectively, and $P$ is the pump power.

Within the framework of the model we find how phase of the comb repetition rate, defined as
\begin{equation} \label{psi}
\Psi(t)= {\rm Arg} \left [ e^{2\pi i (FSR+\delta)\ t} \sum \limits_j {\hat a}_j {\hat a}_{j-1}^\dag \right ],
\end{equation}
is modulated at frequency $f_m$ ($\delta$ corresponds to the deviation of the comb repetition rate from the FSR, occurring due to the presence of TOD as well as the wavelength dependent Q-factor); and how the power of the frequency comb, defined as
\begin{equation} \label{pcomb}
P_{comb}(t)=  \sum \limits_j {\hat a}_j {\hat a}_{j}^\dag,
\end{equation}
is modulated at frequency $f_m$.

The external continuous wave pump is applied to the central mode of the mode group ($j=j_0$). The frequencies of  modes are given by Eq.~(\ref{freq}). The free parameters include the pump amplitude, $f$, the frequency detuning, $\Delta=(\nu-\nu_0)/\gamma$; as well as the modulation depth, $m$, and frequency, $f_m$. We found that once the comb is generated  the Feshbach resonance is observed  for both amplitude (real $m$) and  phase (imaginary $m$) modulation, and at all possible pump powers. For the sake of simplicity only results for amplitude modulation are presented here. We select $f=\sqrt{50}$, $m=0.001$,  $D_2/\gamma=0.05$, and $D_3/\gamma=10^{-4}$.

\begin{figure}[htb]
\centerline{\includegraphics[width=7.5cm]{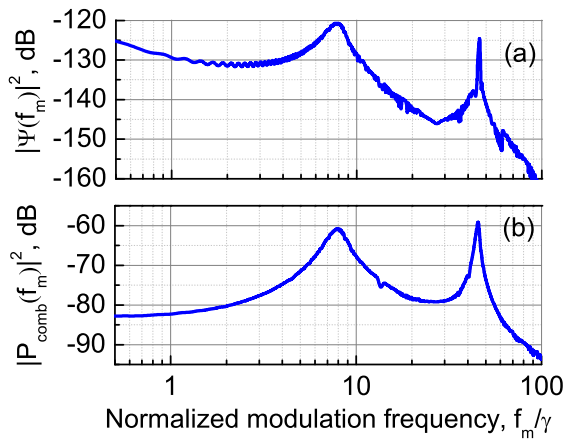}}
\caption{{ Numerically simulated modulation amplitudes of the phase $\Psi(f_m)$, (a), and power $P_{comb}(f_m)$, (b), versus modulation frequency $f_m$. The low frequency peak corresponds to the relaxation oscillation of the Kerr comb that depends on the bandwidth of the resonator modes, and the  high frequency peak corresponds to the frequency detuning $|\Delta|$.  }\label{figure1}}
\end{figure}

Selecting $\Delta=-45$ we simulated the behavior of the fundamentally mode locked frequency comb and found the dependence of the modulation harmonics of parameters defined by Eqs.~(\ref{psi}) and (\ref{pcomb}) on the modulation frequency. In this test we selected the frequency $f_m$, simulated steady state dynamics of the frequency comb for a long enough time, and performed Fourier analysis of $\Psi(t)$ and $P_{comb}(t)$. The frequency comb is at hard excitation regime \cite{matsko12pra}, so we used optimized nonzero initial conditions for the comb harmonics. The result is presented in Fig.~(\ref{figure1}).

The simulation reveals two resonances in the frequency dependence. The smaller frequency resonance, which results from the relaxation oscillation, occurs at approximately $f_m=(2 \pi^3)^{1/2} \gamma$. Its position only weakly depends on TOD and $Q(\nu)$ dependence. The magnitude of the resonance drops as those perturbations decrease. The higher frequency resonance occurs exactly at frequency $\nu_0-\nu$, as the analytical calculation predicts. The position of the resonance does not depend on $\gamma$ and pump power. The magnitude of the resonance drops when TOD is reduced. The shape of the resonance curve for $P_{comb}(f_m)$ dependence is bell-like, with the center at $\nu_0-\nu$. The shape of the resonance curve for $\Psi(f_m)$ is dispersion-like, with the maximum slope position at $\nu_0-\nu$. This resonance corresponds to the Feshbach resonance predicted by the analytical theory.

The increase of $|\Psi(f_m)|$ at small frequencies $f_m$ is an artefact of the simulation resulting from an incomplete compensation of the frequency shift of the comb repetition rate. The Kerr frequency comb repetition rate coincides with the resonator FSR in the ideal case described by Eq.~(\ref{basic1}). Nonzero TOD results in shift of the frequency by parameter $\delta$. We find this parameter numerically and remove it from the phase $\Psi(f_m)$. Since the removal is not very accurate, the phase drifts at small frequencies, which corresponds to the increase of $\Psi(f_m)$ at $\gamma \gg f_m$.

\begin{figure}[htb]
\centerline{\includegraphics[width=7.5cm]{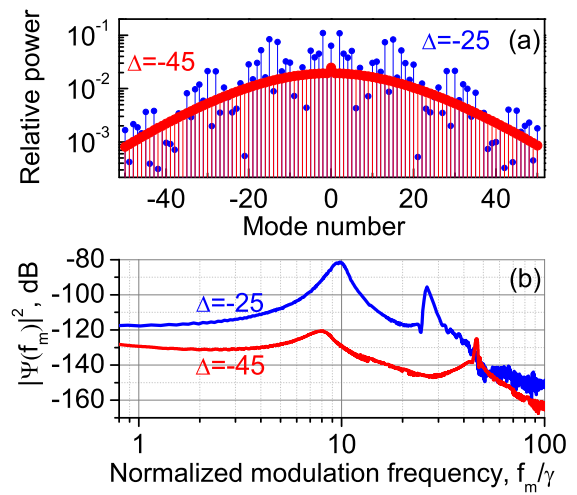}}
\caption{{ Numerically simulated frequency combs for two detuning values $\Delta_0=-25$ (harmonically mode locked comb, blue line) and $\Delta_0=-45$ (fundamentally mode locked comb, red line), (a); and  modulation amplitude of the phase parameter, $\Psi(f_m)$, (b), versus modulation frequency $f_m$.  The low frequency peak in panel (b) corresponds to the relaxation oscillation of the Kerr comb that depends on the bandwidth of the resonator modes, and the  high frequency peak corresponds to the frequency detuning $|\Delta|$. }\label{figure2}}
\end{figure}

The analytical theory is limited in its ability and cannot be used to describe an arbitrary regime of the Kerr frequency comb operation. We utilize numerical simulations to find how the Feshbach resonance behaves in the case of harmonic mode locking. This regime corresponds to generation of multiple pulses in the resonator. Since the harmonic mode locking can be observed more readily at smaller detuning $|\Delta|$, we selected $\Delta = -25$. The results of the simulation are shown in Fig.~(\ref{figure2}).

We found that the relaxation oscillation peak increases in magnitude and shifts to higher frequency, corresponding to a smaller detuning, for the harmonically mode locked comb, as compared with the relaxation oscillation peak of the fundamentally mode locked frequency comb. The Feshbach resonance position of the harmonically mode locked comb corresponds to the detuning, $\nu_0-\nu$. The resonance broadens and increases in amplitude as compared with the resonance observed with the fundamentally mode locked frequency comb.

We performed simulations for a variety of parameters and found that both the relaxation oscillation and Feshbach resonances exist for any mode locked frequency comb in a non-ideal resonator. The importance of this effect is that it results in the transfer of the  amplitude and phase noise of the pump laser to the comb repetition rate as well as comb amplitude noise. This effect is similar to the noise transfer predicted for the resonant Brillouin lasers \cite{matsko12oe} and mode locked lasers \cite{lee14prl}.

In conclusion, we found theoretically the presence of a Feshbach resonance in a resonant Kerr frequency comb oscillator. The existence of the resonance was proven using an analytical approach based on approximate solution of the Lugiato-Lefever  equation, and was confirmed via numerical simulation of an equivalent set of ordinary differential equations. The physical processes discussed in this Letter are useful for an in-situ measurement of the resonator bandwidth and detuning via the relaxation oscillation resonance and the Feshbach resonance frequencies.

The authors acknowledge stimulating discussions with Prof. T. R. Schibli as well as support from Defense Sciences Office of Defense Advanced Research Projects Agency under contract No. W911QX-12-C-0067.

\end{document}